\documentclass[showpacs,twocolumn,secnumarabic,amssymb, nobibnotes, aps, prl]{revtex4-1}
\usepackage{graphicx}
\usepackage{amsmath}
\usepackage{xspace}

\newtheorem{Thm}{Theorem}

\def\QEDopen{{\setlength{\fboxsep}{0pt}\setlength{\fboxrule}{0.2pt}\fbox{\rule[0pt]{0pt}{1.3ex}\rule[0pt]{1.3ex}{0pt}}}}
\def\QED{\QEDopen}

\def\endproof{\hspace*{\fill}~\QED\par\endtrivlist\unskip}

\newcommand\ket[1]{| #1 \rangle}

\begin{document}

\title{A New Paradigm for Quantum Nonlocality}%

\author{Yang D. Li}%
\email[Email: ]{danielliy@gmail.com}
\affiliation{Advanced Digital Sciences Center\\
University of Illinois at Urbana-Champaign}
\date{\today}
\begin{abstract}
Bell's theorem \cite{Bel1964, CHSH1969} basically states that local hidden variable theory cannot predict the correlations produced by quantum mechanics. It is based on the assumption that Alice and Bob can choose measurements from a measurement set containing more than one element. We establish a new paradigm that departs from Bell's paradigm by assuming that there are no choices for Alice and Bob and that the measurements Alice and Bob will make are fixed from the start. We include a process of quantum computation in our model. To the best of our knowledge, we are the first to connect quantum computation and nonlocality, the two faces of entanglement.
\end{abstract}
\pacs{03.65.Ud, 03.65.Ta, 03.67.Mn}
\maketitle

The charm and beauty of quantum mechanics is grounded on its counterintuitiveness to a great extent. And one of the most counterintuitive results ever in history is Bell's theorem \cite{Bel1964, CHSH1969}, implying that quantum mechanics violates either locality or counterfactual definiteness. A simple and informal restatement of Bell's theorem is that local hidden variable theory cannot reproduce all of the predictions of quantum mechanics.

We think about the power of entanglement and the simulation of entanglement from another prospective. In Bell's theorem, the measurements are versatile and can be chosen be with respect to various bases. Alice may choose to measure her state according to basis with an angle $\alpha$ relative to the standard basis; while Bob may choose another basis, which is $\beta$ relative to the standard basis, for measurement. What is the case if we fix the measurement? Is it now possible to simulate ``quantum mechanics" with local hidden variables?

Next we shall make our model more formal and more concrete. The model is roughly illustrated by Figure $1$.

\begin{figure}
\begin{center}
\includegraphics[angle=0, width=0.5\textwidth]{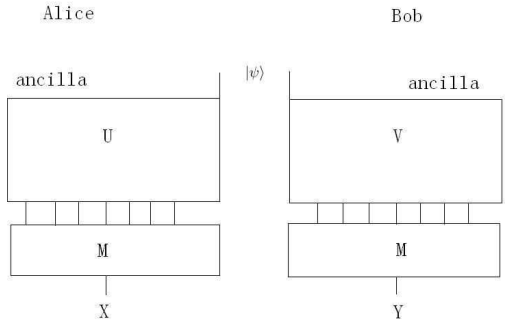}
\caption{\label{Figure 1}A New Paradigm}
\end{center}
\end{figure}

In Figure $1$, $\ket{\psi}$ represents an entangled state of size $2Q$ ($Q \le n$) qubits, where the first $Q$ qubits are owned by Alice and the second $Q$ qubits belong to Bob. We may also add some ancilla qubits to make sure that both Alice and Bob have $n$ qubits. The state in the beginning is $\phi_0$. Alice applies unitary operation $U$ and Bob applies unitary operation $V$. The state becomes $\phi_1 = (U \otimes V)\phi_0$. $U$ and $V$ are fixed for Alice and Bob. The role of $U$ and $V$ can be seen as a step of quantum computation to make the quantum correlation harder to simulate classically.

Then Alice and Bob both apply the measurement $M$, which is fixed to be with respect to the standard basis. $M$ is assumed to be fixed from the start. In the end, they output a correlation $(X,Y)$ according to results of the measurement, where $X$ and $Y$ are random variables taking values in $\{0, 1\}^n$. Here, we use $\{0, 1\}^n$ to denote a set containing $2^n$ binary strings of length $n$. For instance, if $n = 2$, then $\{0, 1\}^2$ is a set containing $2^2 = 4$ binary strings: $\{00, 01, 10, 11\}$. Now it is easy to understand $X$ and $Y$. If $n = 1$, then both $X$ and $Y$ have two possible outcomes, i.e., $0$ and $1$. $X$ equals to $0$ with some probability $p$ ($0 \le p \le 1$) and equals to $1$ with the remaining probability $1 - p$. $Y$ takes value $0$ with probability $q$ ($0 \le q \le 1$) and takes value $1$ with probability $1 - q$. $X$ and $Y$ are correlated in the sense that their distributions are not independent. Suppose the distribution of $(X, Y)$ is $P_r$, and we want to approximately simulate $P_r$ using local hidden variables. Note that $P_r$ can be treated as a matrix, namely
\begin{equation}
(P_r)_{xy} = Prob\{X = x, Y = y\},
\end{equation}
for all $x, y \in \{0, 1\}^n$.

There are two clear differences of our paradigm from Bell's. First, we add a process of quantum computation; namely, Alice and Bob apply unitary operations on the initial quantum state. In Bell's setting, there is no such consideration: Alice and Bob measure the quantum state directly. Secondly, in Bell's model, Alice and Bob have the freedom to choose their measurement; while here the measurement of Alice and Bob is prescribed from the start. As we will see, the key feature of the new model is that there is no so-called free-will and space-like separation between Alice and Bob's measurement choices. Thus it is possible to build a local model consistent with quantum correlation. But we show that it is still very hard to simulate quantum correlation and there could be some pathology.

For the classical simulation, Alice and Bob initially share some random bits (shared randomness, public coins, or local hidden variables). We denote the shared random variable as $Z$. Alice and Bob also have some private random bits, denoted as $r_A$ and $r_B$ respectively. They use $Z$, $r_A$ and $r_B$ to generate a correlation $(X', Y')$, such that $X' = f_A(Z, r_A)$ and $Y' = f_B(Z, r_B)$. $X'$ and $Y'$ are random variables taking values in $\{0, 1\}^n$. Suppose that the probability distribution of $(X', Y')$ is $P_c$, we want to make sure that $P_c$ is close to $P_r$. If $P_c$ and $P_r$ are close enough, then we succeed in simulating quantum correlations using local hidden variables.

Now the question is: how to measure the distance between two probability distributions $P_c$ and $P_r$? There are usually two ways. One is to allow some multiplicative error:
\begin{equation}
(1 - \beta)(P_r)_{xy} \le (P_c)_{xy}\le (1+\beta)(P_r)_{xy},
\end{equation}
for all $x, y \in \{0, 1\}^n$, where $\beta$ is some constant satisfying $0 \le \beta < 1$. In this case, we say that $P_c$ and $P_r$ are $\beta$-close. The other one is to allow some additive error, which is used more frequently in practice. For two probability distributions $D_1$ and $D_2$ over a space $S$ (in our case, the space $S$ is $\{0, 1\}^n \times \{0, 1\}^n$), the \emph{variational distance} between them is
\begin{equation}
||D_1 - D_2||_1 = \sum_{s \in S} |D_1(s) - D_2(s)|.
\end{equation}
If $P_c$ and $P_r$ satisfy $||P_c - P_r||_1 \le \epsilon$, then we say that the variational distance between $P_c$ and $P_r$ is at most $\epsilon$, for some constant $\epsilon > 0$.

For the multiplicative-error case, we have the following theorem.
\begin{Thm}\label{MT1}
There is an entangled state $\ket{\psi}$ with $Q = 1$ and some $U$ and $V$, such that for any given constant $0 \le \beta < 1$, at least $\log_2(n)$ shared random bits are needed to produce $P_c$, where $P_c$ and $P_r$ are $\beta$-close.
\end{Thm}
For the additive-error case, we show the following result.
\begin{Thm}\label{MT2}
There is an entangled state $\ket{\psi}$ with $Q = O(\log_2(n))$, and some $U$ and $V$, such that for any given constant $\epsilon > 0$, at least $\Omega(\sqrt{n})$ shared random bits are needed to generate $P_c$, where the variational distance between $P_c$ and $P_r$ is at most $\epsilon$.
\end{Thm}

In Theorem \ref{MT2}, the $O(\log_2(n))$ notation ($O$ is called \emph{big-Oh} notation) means that $Q$ is bounded above by $\log_2(n)$ up to some constant and $\Omega(\sqrt{n})$ ($\Omega$ is called \emph{big-Omega} notation) notation means that the amount of shared random bits is bounded below by $\sqrt{n}$ up to some constant. In other words, $Q$ is at most $c_1\log_2(n)$ for some constant $c_1 > 0$ and the amount of shared random bits needed is at least $c_2\sqrt{n}$, where $c_2 > 0$ is some constant.

Theorem \ref{MT1} is an ``$\log_2(n)$ vs. $1$" (super-exponential) separation and Theorem \ref{MT2} is an ``$\Omega(\sqrt{n})$ vs. $O(\log_2(n))$" (exponential) separation. Hence, we have shown that quantum entanglement is still much more powerful than classical shared randomness, even when we fix the measurement and consider the approximate simulation. This result significantly expands our knowledge on the power of entanglement and helps us better understand the principles of
quantum entanglement.

Theorem $\ref{MT1}$ says that finite amount of local hidden variables cannot account for quantum correlations even when the measurements Alice and Bob will make are prescribed from the start. For any finite amount of local hidden variables, there are always quantum correlations it cannot explain. This is due to the fact that $n$ can be arbitrarily large while $Q = 1$ is fixed.

Theorem \ref{MT2} is very interesting and subtle: local hidden variables cannot account for quantum correlations even when the measurements Alice and Bob will make are prescribed from the start, which is delicate because it is not true for any finite set of measurements. That is, Alice and Bob share some number $n$ of pairs and for any finite $n$, local hidden variables \textit{can} produce the results of their measurements. Local hidden variables fail, not at any particular value of $n$, but in the limit as $n$ tends to infinity, because the number of local hidden variables needed to account for the results grows exponentially. Maybe we can say that the problem is that there is no thermodynamic limit. That is to say, the number of hidden variables per pair shared by Alice and Bob is not an intensive quantity. Our argument is not as convincing as Bell's, but it goes beyond Bell's theorem in the sense that it shows that if hidden variables explain quantum correlations, then there is some pathology in the explanation.

We give a short overview of the proofs of Theorem \ref{MT1} and \ref{MT2}. Theorem \ref{MT1} is relatively easy to prove. We set the initial state to be the well-known Bell state and go on to compute $P_r$. We construct a $P_r$ that is very hard to simulate using local hidden variables by setting a proper $U$ and $V$, and thus complete the proof. The proof of theorem \ref{MT2} is much more complicated.
We first define a specific distribution $P_u$. Based on $P_u$, we set the initial state, $U$, $V$ and make sure that $P_r$, the distribution after measurement, is very close to $P_u$. So if we want to guarantee that $P_c$, the distribution generated classically, is close to $P_r$, then we have to assure that $P_c$ is also close to $P_u$. And we show that the classical simulation of $P_u$ is hard and thus complete the proof.

Although the mathematics in the proof of Theorem \ref{MT2} is very complicated, the intuitive idea is very simple: to approximately generate a $2n$-bit correlated classical information (here it means $P_u$), roughly $\log_2(n)$ shared qubits are enough (here it is the initial quantum state from which we get $P_r$), but classically we need at least (roughly) $\sqrt{n}$ correlated bits (here this is the shared random bits from which we generate $P_c$). This exponential separation is inherent. We can imagine the following situation. Suppose that Alice and Bob share some number $n$ of pairs. If the initial state shared by Alice and Bob is a product state, then the probability distribution of their joint results would be simply the product of the probability distributions of their separate results and the amount of information needed to specify the probability distribution of their joint results would automatically be proportional to the number $n$ of shared pairs. But if the initial state is an entangled state, like in our case, the number of terms in its Schmidt decomposition can be exponential in $n$, and the amount of information needed is not necessarily proportional to $n$.

{\em Proof of Theorem \ref{MT1}: } We set $\ket{\psi}$ to be the Bell state
\begin{equation}
\ket{\psi} = \frac{\ket{00}+\ket{11}}{\sqrt{2}},
\end{equation}
where the first qubit is owned by Alice while the second belongs to Bob. So $Q = 1$. We also fill some ancilla qubits. The state in the beginning is
\begin{equation}
\phi_0 = \ket{0^{n - 1}} \ket{\psi} \ket{0^{n - 1}}.
\end{equation}
Then Alice applies unitary operation $U$ and Bob applies unitary operation $V$. $U$ and $V$ are unitary matrices of size $2^n \times 2^n$. The state becomes
\begin{equation}
\phi_1 = \frac{1}{\sqrt{2}} (u_0 \otimes v_0 + u_1 \otimes v_1),
\end{equation}
where $u_0$ is the first column of $U$, $v_0$ is the first column of $V$, $u_1$ is the second column of $U$, and $v_1$ is the $(2^{n - 1} + 1)$-th column of $V$. Suppose $N = 2^n$. After the measurement $M$, there are $N \times N$ possibilities, and
\begin{equation}
(P_r)_{xy} =  \frac{1}{2}|u_0(x)v_0(y) + u_1(x)v_1(y)|^2,
\end{equation}
for all $x, y\in \{0,1\}^n$.

Next we set the proper $U$ and $V$. Let $\{c_x : x \in \{0, 1\}^n\}$ be a set of $N$ distinct real numbers. If we take $\{c_x : x \in \{0, 1\}^n\}$ to be proper values \cite{SM} satisfying
\begin{equation}
\sqrt{\sum_{1\le x < y \le n}(c_y - c_x)^2} = \sqrt{1/2},
\end{equation}
and let $v_0 = \bar{u}_0$ and $v_1 = - \bar{u}_1$, we have
\begin{equation}
P_r = [(c_y - c_x)^2]_{xy} = [\frac{1}{2}|u_0(x)\bar{u}_0(y) - u_1(x)\bar{u}_1(y)|^2]_{xy}.
\end{equation}
Here, $\bar{v}$ is the conjugate vector of $v$. Therefore, the diagonal entries of such a $P_r$ are $0$ and its off-diagonal entries are non-zero.

Suppose we want to use shared random bits to generate a $P_c$ that is $\beta$-close to $P_r$. Any $P_c$ has to satisfy
\begin{equation}
(1 - \beta)(P_r)_{xy} \le (P_c)_{xy}\le (1+\beta)(P_r)_{xy},
\end{equation}
for all $x, y \in \{0, 1\}^n$. Thus, $P_c$ also has the property that its diagonal entries are $0$ and its off-diagonal entries are non-zero.
Suppose that in the beginning Alice and Bob share a random variable $Z$, whose sample space is $S$. The size of $S$ is bounded below by $\log_2(N)$ \cite{SM}.
Consequently,
\begin{equation}
\log_2(|S|) \ge \log_2(n),
\end{equation}
implying that we need at least $\log_2(n)$ bits of shared random bits to approximately simulate $P_r$. This completes the proof of Theorem \ref{MT1}. \endproof

{\em Proof of Theorem \ref{MT2}: } First, we want to define a probability distribution $P_u$ over the space $\{0, 1\}^n \times \{0, 1\}^n$. Suppose $x \in \{0, 1\}^n$. $x$ is a $n$-bit binary strings, and we use $|x|$ to denote the \emph{cardinality} of $x$, namely the number of $1$'s in $x$. For example, if $n = 8$ and $x = 00111001$, then the cardinality of $x$ is $|x| = 4$. Also suppose $y \in \{0, 1\}^n$. If $x_i \wedge y_i = 0$, for all $i \in \{1, 2, \ldots, n\}$, then we say that $x$ and $y$ are disjoint. For instance, if $n = 4$, $x = 1100$ and $y = 0011$, then $x$ and $y$ are disjoint. Without loss of generality, we assume that $n$ is a square number, such as $1, 4, 9, 16$ and so on, and that $N_1 = {{n-\sqrt{n}} \choose \sqrt{n}}$, $N_2 = {{n\choose \sqrt{n}}}$. Here the notation ${n \choose k}$ means the binomial coefficient, namely the number of ways of selecting $k$ things out of a group of $n$ elements. We define a probability distribution $P_u$ (in the form of a matrix) to be
\begin{equation}
(P_u)_{xy} = \frac{1}{N_1N_2}
\end{equation}
if and only if $|x| = |y| = \sqrt{n}$ and $x$ and $y$ are disjoint; otherwise $(P_u)_{xy} = 0$. In words, $P_u$ is a uniform distribution over all the disjoint $x$ and $y$ whose cardinality are both $\sqrt{n}$. Another matrix $M_u$ can be defined based on $P_u$ :
\begin{equation}
(M_u)_{xy} = \sqrt{(P_u)_{xy}},
\end{equation}
for all $x, y \in \{0, 1\}^n$. By spectral decomposition, there exists a unitary matrix $U_1$ such that
\begin{equation}
M_u = U_1D_uU_1^{\dagger},
\end{equation}
where $D_u$ is a diagonal matrix consisting of the spectrum $\lambda_0, \lambda_1, \ldots, \lambda_{2^n - 1}$ ($\lambda_0 \ge \lambda_1 \ge \cdots \ge \lambda_{2^n - 1}$).

Define $K$ to be
\begin{equation}
K = \min\{k: \sum_{i \le k} \lambda_i^2 \ge 1 - \frac{\epsilon^2}{8}\},
\end{equation}
and set $Q$ to be $\lceil \log_2(K+1) \rceil$, $U$ to be $U_1$, and $V$ to be $\overline{U_1}$. Then in our model,
\begin{equation}
\psi = \frac{1}{N'}\sum_{i = 0}^{K} \lambda_{i} \ket{i, i},
\end{equation}
\begin{equation}
\phi_0 = \frac{1}{N'}\sum_{i = 0}^{K} \lambda_{i} \ket{00\cdots0i, 00\cdots0i},
\end{equation}
and
\begin{equation}
\phi_1 = (U_1 \otimes \overline{U_1})\phi_0,
\end{equation}
where $N'$ is a normalization factor and $\sqrt{1 - \frac{\epsilon^2}{8}} \le N' \le 1$. It is not hard to verify that
\begin{equation}
(P_r)_{xy} = \frac{1}{N'^2}|(U_1D_rU_1^{\dagger})_{xy}|^2,
\end{equation}
where $D_r = \text{diag}\{\lambda_0, \lambda_1, \ldots , \lambda_{K}, 0, 0, \ldots, 0\}$. The variational distance of $P_u$ and $P_r$ is at most $\epsilon$, and $Q$ is at most (roughly) $\log_2(n)$, namely $Q = O(\log_2(n))$ \cite{SM}.

Since $||P_u - P_r||_1 \le \epsilon$, if we want to ensure $||P_c - P_r||_1 \le \epsilon$, we have to guarantee $||P_c - P_u||_1 \le 2\epsilon$; otherwise
\begin{equation}
||P_c - P_r||_1 \ge ||P_c - P_u||_1 - ||P_u - P_r||_1 \ge \epsilon,
\end{equation}
which does not satisfy our aim. Next we shall see the amount of classical local hidden variables needed to produce $P_c$, where $||P_c - P_u||_1 \le 2\epsilon$. Suppose in the beginning Alice and Bob share a random variable $Z$, whose sample space is $S$. Conditional on $Z$, $X'$ and $Y'$ are independent. We denote the distribution of $(X', Y')$ under $Z = z$ to be $D_z$,
which is a product distribution. Thus,
\begin{equation}
P_c = \sum_{z \in S}Prob\{Z = z\}D_z.
\end{equation}
The size of $S$ is lower bounded by $2^{\Omega(\sqrt{n})}$ \cite{SM}. Thus, the amount of shared random bits needed is at least $\log_2(S) = \Omega(\sqrt{n})$.  This completes the proof of Theorem \ref{MT2}. \endproof

{\em Related Work: }The idea of a great deal of papers (\cite{BCT1999, BCvD2001, BT2003, TB2003, RT2009}, etc.) is that local hidden variables augmented by communication could reproduce the results of quantum entanglement. Quantum entanglement also has plenty of applications in areas such as quantum teleportation \cite{BBCJPW1993}, superdense coding \cite{BW1992} and quantum cryptography \cite{BB1984}. In our previous work \cite{Li2011}, it is shown that at least $n$ bits of local hidden variables are needed to exactly simulate the correlation generated from a $2$-qubit Bell state. This is a ``$n$ vs. $1$" (super-exponential) separation, and is very interesting theoretically. But it has a fatal shortcoming in the sense that it cannot be experimentally validated. There is simply no effective way to tell if two probability distributions are exactly the same. As a result, in this paper we focus on the approximate simulation, which is intriguing in theory, and at the same time feasible in experiment. \cite{Zha2010} is in a game theoretic setting and the players there have some free will. In addition, the result in \cite{Zha2010} deals with exact simulation instead of approximate simulation, and is wrong \cite{Li2011b}. \cite{Win2005} is in an information theoretic setting, and players there have free will. Moreover, \cite{Win2005} only discusses simulating some distribution and does not discuss simulating the distribution generated from quantum measurement.

{\em Concluding Remarks: }In a nutshell, we have established a new paradigm for quantum nonlocality by restricting the set of measurements and using quantum computation. We also obtained two separation results under the new model to show the hardness of classically explaining quantum mechanics using local hidden variables. Our work is seminal in at least two aspects. First, the discussions of Bell's theorem are always complicated by counterfactual argumentation. In our paper, there are no counterfactuals, so we have a fresh approach to and a new look at Bell's theorem. Secondly, quantum information theory relies on entanglement in two quite different contexts: as a resource for quantum computation and as a source for nonlocal correlations among different parties. It is strange and not understood that nonlocality is crucially linked with entanglement in the second context but not in the first. Quantum computation and nonlocality are two faces of entanglement that we do not usually connect, and in this paper we create a direct interface between quantum computation and nonlocality. To the best of our knowledge, we are the first to do so.

We are grateful to the anonymous referees for their insightful and helpful comments.

\bibliography{Appro_Bell}
\end{document}